\documentclass[preprintnumbers,amsmath,amssymbm,prd]{revtex4}
\usepackage{epsfig}
\usepackage{graphicx}

\begin{document}
\title{A test of the conjectured critical black-hole formation -- null geodesic correspondence: 
The case of self-gravitating scalar fields}
\author{Shahar Hod}
\affiliation{The Ruppin Academic Center, Emeq Hefer 40250, Israel}
\affiliation{ }
\affiliation{The Hadassah Institute, Jerusalem 91010, Israel}
\date{\today}

\begin{abstract}
\ \ \ It has recently been conjectured [A. Ianniccari {\it et al.}, Phys. Rev. Lett. {\bf 133}, 081401 (2024)] 
that there exists a correspondence between the critical threshold of 
black-hole formation and the stability properties of null circular geodesics 
in the curved spacetime of the collapsing matter configuration. 
In the present compact paper we provide a non-trivial test of this intriguing conjecture. 
In particular, using analytical techniques we study the physical and mathematical properties of 
self-gravitating scalar field configurations that possess marginally-stable (degenerate) null circular geodesics. 
We reveal the interesting fact that the {\it analytically} calculated critical compactness 
parameter ${\cal C}^{\text{analytical}}\equiv{\text{max}_r}\{m(r)/r\}=6/25$, which 
signals the appearance of the first (marginally-stable) null circular geodesic in the curved spacetime of the self-gravitating scalar fields, agrees 
quite well (to within $\sim10\%$) with the exact compactness 
parameter ${\cal C}^{\text{numerical}}\equiv\text{max}_t\{\text{max}_r\{m(r)/r\}\}\simeq0.265$ which 
is computed {\it numerically} using fully non-linear numerical simulations of the gravitational collapse of scalar fields at 
the threshold of black-hole formation 
[here $m(r)$ is the gravitational mass contained within a sphere of radius $r$]. 
\end{abstract}
\bigskip
\maketitle

\section{Introduction}

Curved spacetimes of highly compact matter configurations 
may possess closed light rings (null circular geodesics) \cite{Bar,Chan,ShTe} on
which massless particles can orbit the central compact object. 
These null geodesics are of fundamental physical importance in theoretical and observational studies 
of highly curved spacetimes (see 
\cite{Bar,Chan,ShTe,Pod,Ame,Ste,Goe,Mas,Dol,Dec,Hodf,Hodt1,Hodt2,Zde1,Hodcom}
and references therein).

In particular, it has been shown that the optical appearance of a highly compact collapsing star is 
determined by the physical properties of its external light rings \cite{Pod,Ame}. 
Likewise, the physically important phenomenon of strong gravitational lensing 
is closely related to the existence of closed null circular geodesics in the 
curved spacetimes of the central compact objects \cite{Ste}. 
In addition, it is well established that the eikonal quasinormal resonance spectra that 
characterize the relaxation dynamics of highly compact self-gravitating objects (black holes and spatially regular ultra-compact objects that 
possess light rings) are determined by the physical properties of these closed null circular geodesics (see 
\cite{Goe,Mas,Dol,Dec,Hodf} and references therein). 

Intriguingly, Ianniccari {\it et al.} \cite{IIKPR} have recently provided evidence that null circular geodesics 
may also be related to the formation of primordial black holes. 
In particular, it has been demonstrated in the physically important work \cite{IIKPR} that there is a correspondence 
between the critical threshold of primordial black-hole formation and the appearance of the 
first (marginally-stable) null circular geodesic in the curved spacetime of the collapsing matter configuration. 

The main goal of the present compact paper is to test, using {\it analytical} techniques, the 
validity of the black-hole formation -- null geodesic correspondence proposed in the physically 
interesting work \cite{IIKPR}. To this end, we shall analyze the physical and mathematical 
properties of self-gravitating scalar field configurations that possess light rings. 
In particular, we shall determine the critical value, 
\begin{equation}\label{Eq1}
{\cal C}^{\text{analytical}}\equiv\text{max}_r\Big\{{{m(r)}\over{r}}\Big\}\  ,
\end{equation}
of the dimensionless compactness parameter which signals the appearance of the first null circular geodesic 
in the curved spacetime of the self-gravitating field configuration. 

Interestingly, below we shall reveal the fact that the {\it analytically} derived value 
${\cal C}^{\text{analytical}}$ of the dimensionless compactness parameter, which signals 
the critical formation of a marginally-stable null circular geodesic in the curved spacetime of the self-gravitating 
scalar configuration, agrees quite well (to within $\sim10\%$) with the corresponding exact 
value ${\cal C}^{\text{numerical}}$ of the compactness parameter 
as determined from fully non-linear {\it numerical} simulations \cite{Chop1} 
of the gravitational collapse of scalar fields at the critical threshold of black-hole formation (see 
also \cite{HodPir1,HodPir2} and references therein). 

\section{Description of the system}

We consider self-gravitating scalar field configurations which are characterized by the spherically 
symmetric curved line element \cite{Chan,Noteunits}
\begin{equation}\label{Eq2}
ds^2=-\alpha^2(r,t) dt^2 +a^2(r,t)dr^2+r^2(d\theta^2 +\sin^2\theta d\phi^2)\  .
\end{equation}
The Einstein equations yield the dimensionless functional relations \cite{Chop1,Notepm}
\begin{equation}\label{Eq3}
{{r\alpha'}\over{\alpha}}-{{ra'}\over{a}}+1-a^2=0\
\end{equation}
and
\begin{equation}\label{Eq4}
a(r)=\Big[{1-{{2m(r)}\over{r}}}\Big]^{-1/2}\
\end{equation}
for the metric functions, where \cite{Noterho}
\begin{equation}\label{Eq5}
m(r)=\int_0^r 4\pi x^2\rho(x)dx\
\end{equation}
is the gravitational mass contained within a sphere of radius $r$. 

It is convenient to define the dimensionless compactness function 
\begin{equation}\label{Eq6}
{\cal C}(r)={{m(r)}\over{r}}\  ,
\end{equation}
which is characterized by the physically motivated boundary condition 
\begin{equation}\label{Eq7}
{\cal C}(r=0)=0\ 
\end{equation}
of a regular origin. 

\section{Testing the conjectured critical black-hole formation -- null geodesic correspondence}

In the present section we shall test the validity of the physically intriguing correspondence suggested 
in \cite{IIKPR} between the threshold of black-hole formation and the appearance of the 
first (marginally-stable) circular geodesic in the corresponding curved spacetime of the self-gravitating matter configuration. 

Following \cite{IIKPR} we shall consider a physical situation in which the curved spacetime 
is temporarily stationary. 
It is interesting to note that this assumption is particularly suitable for the present model 
of a self-gravitating scalar field whose critical solution at the threshold of black-hole formation has a discrete self-similar 
character \cite{Chop1,HodPir1,HodPir2}. 
Thus, the critical solution of the scalar field model at the threshold of black-hole formation 
is characterized by a discrete set of stationary times $\{t_n\}^{n=\infty}_{n=1}$ for 
which physically measurable quantities, like the dimensionless 
compactness parameter ${\cal C}_{\text{max}}(t)\equiv\text{max}_r\{{\cal C}(r)\}$, are temporarily stationary. 
In particular, below we shall use analytical techniques in order to estimate the maximum value,
\begin{equation}\label{Eq8}
{\cal C}^*_{\text{max}}\equiv\text{max}_t\{\text{max}_r\{{\cal C}(r,t)\}\}\  ,
\end{equation}
of the compactness parameter which is defined by the relation 
\begin{equation}\label{Eq9}
{{d{\cal C}_{\text{max}}}\over{dt}}=0\ \ \ \ \text{for}\ \ \ \ t\in\{t_n\}^{n=\infty}_{n=1}\  .
\end{equation}

The condition for the existence of light ring(s) in 
the curved spacetime is given by the compact functional relation \cite{IIKPR,Hodev1,Cun,Hodev2}
\begin{equation}\label{Eq10}
{\cal F}(r)\equiv{{r\alpha'}\over{\alpha}}-1=0\ \ \ \ \text{for}\ \ \ \ r=r_{\text{c}}\  .
\end{equation}
It is well established \cite{Hodev1,Cun,Hodev2} that spatially regular curved spacetimes generally possess 
an {\it even} (or zero) number of null circular geodesics (closed light rings). 
In particular, the first appearance of a marginally-stable light ring in the 
curved spacetime is characterized by the degenerate functional relation \cite{IIKPR,Hodev1,Cun,Hodev2}
\begin{equation}\label{Eq11}
{\cal F}(r_{\text{c}})={\cal F}'(r_{\text{c}})=0\  .
\end{equation}
From Eqs. (\ref{Eq10}) and (\ref{Eq11}) one obtains the characteristic relation 
\begin{equation}\label{Eq12}
\alpha''=0\ \ \ \ \text{for}\ \ \ \ r=r^*_{\text{c}}\
\end{equation}
for the radial location, $r=r^*_{\text{c}}$, of the marginally-stable (degenerate) 
null circular geodesic in the curved spacetime. 

Taking cognizance of Eqs. (\ref{Eq3}), (\ref{Eq4}), (\ref{Eq10}), and (\ref{Eq12}) one obtains 
the two coupled equations
\begin{equation}\label{Eq13}
ra'-(2-a^2)a=0\ \ \ \ \text{for}\ \ \ \ r=r^*_{\text{c}}\
\end{equation}
and
\begin{equation}\label{Eq14}
(a'+ra'')a-ra'^2+2a^3a'=0\ \ \ \ \text{for}\ \ \ \ r=r^*_{\text{c}}\  ,
\end{equation}
which yield the functional relations [see Eqs. (\ref{Eq1}) and (\ref{Eq4})] \cite{Noteaaa}
\begin{equation}\label{Eq15}
4{\cal C}_{\text{c}}+r^*_{\text{c}}{\cal C}'_{\text{c}}=1\
\end{equation}
and
\begin{equation}\label{Eq16}
20{\cal C}_{\text{c}}-{r_{\text{c}}^*}^2{\cal C}''_{\text{c}}=5\
\end{equation}
for the marginally-stable null circular geodesic. 
[The subscript $c$ means that the physical quantities are 
evaluated at the critical radius $r=r^*_{\text{c}}$ of the marginally-stable (degenerate) null circular 
geodesic]. 

In order to determine analytically the values of the dimensionless physical 
quantities $\{{\cal C}_{\text{c}},r^*_{\text{c}}{\cal C}'_{\text{c}},{r_{\text{c}}^*}^2{\cal C}''_{\text{c}}\}$ we 
shall use the functional expansion 
\begin{equation}\label{Eq17}
{\cal C}(r\simeq r^*_{\text{c}})={\cal C}_{\text{c}}+{\cal C}'_{\text{c}}\cdot(r-r^*_{\text{c}})+
{1\over2}{\cal C}''_{\text{c}}\cdot(r-r^*_{\text{c}})^2+O\{[(r-r^*_{\text{c}})/r^*_{\text{c}}]^3\}\
\end{equation}
of the compactness function. 
Taking cognizance of Eqs. (\ref{Eq7}), (\ref{Eq15}), (\ref{Eq16}), and (\ref{Eq17}) one 
finds the dimensionless relations \cite{Notebrd}
\begin{equation}\label{Eq18}
{\cal C}_{\text{c}}={{7}\over{30}}\ \ \ \ ; \ \ \ \ r^*_{\text{c}}{\cal C}'_{\text{c}}={{1}\over{15}}\ \ \ \ ; 
\ \ \ \ {r_{\text{c}}^*}^2{\cal C}''_{\text{c}}=-{1\over3}\  .
\end{equation}

The peak (maximum value) of the characteristic compactness function is determined by the gradient relation 
\begin{equation}\label{Eq19}
{\cal C}'(r=r_{\text{p}})=0\
\end{equation}
which, using Eqs. (\ref{Eq17}) and (\ref{Eq18}), yields the dimensionless relations
 \begin{equation}\label{Eq20}
{{r_{\text{p}}-r^*_{\text{c}}}\over{r^*_{\text{c}}}}={1\over5}\
\end{equation}
and 
\begin{equation}\label{Eq21}
{\cal C}^{\text{analytical}}\equiv{\cal C}(r=r_{\text{p}})={{6}\over{25}}\  .
\end{equation}

\section{Summary and discussion}

Motivated by the important results recently presented in \cite{IIKPR}, 
which provide compelling evidence for an interesting relation between the critical threshold of 
black-hole formation and the stability properties of closed light rings (null circular geodesics) 
in the curved spacetime of the collapsing matter configurations, 
we have analyzed the physical and mathematical properties of self-gravitating scalar field configurations. 

In particular, using analytical techniques we have proved that the critical compactness parameter, 
which signals the appearance of the first (marginally-stable) light ring in the curved spacetime of the field configuration, 
is given by the compact relation [see Eqs. (\ref{Eq1}), (\ref{Eq6}), and (\ref{Eq21})]
\begin{equation}\label{Eq22}
{\cal C}^{\text{analytical}}\equiv\text{max}_r\Big\{{{m(r)}\over{r}}\Big\}={{6}\over{25}}\  .
\end{equation}

In order to test the validity of the conjectured critical black-hole formation -- null geodesic 
correspondence suggested in \cite{IIKPR}, one should compare the {\it analytically} derived value (\ref{Eq22}) of 
the critical compactness parameter with the corresponding exact ({\it numerically} computed) 
value ${\cal C}^{\text{numerical}}_{\text{max}}$ of the compactness parameter 
as determined from fully non-linear numerical simulations 
of the gravitational collapse of scalar fields at the critical threshold of black-hole 
formation \cite{Chop1,HodPir1,HodPir2}. 
In particular, from Figs. $3$ and $4$ of \cite{HodPir1} one finds the numerically computed 
value [see Eq. (\ref{Eq8})] 
\begin{equation}\label{Eq23}
{\cal C}^{\text{numerical}}_{\text{max}}\simeq0.265\  .
\end{equation}

From Eqs. (\ref{Eq22}) and (\ref{Eq23}) one learns that the {\it analytically} determined value of the critical 
compactness parameter, whose derivation in the present compact paper 
is based on the conjectured black-hole formation -- null geodesic 
correspondence of \cite{IIKPR}, agrees to within $\sim10\%$ with the corresponding exact value of the 
compactness parameter as determined {\it numerically} \cite{Chop1,HodPir1,HodPir2} using fully non-linear simulations of the collapse of 
self-gravitating scalar fields at the threshold of black-hole formation.

Thus, our analytical results indicate that, in accord with the physically interesting conjecture made in \cite{IIKPR}, 
there may be a non-trivial relation between the critical threshold of black-hole formation and the 
appearance of the first light ring (marginally-stable null circular geodesic) 
in the curved spacetime of the collapsing matter configuration. 
  

\bigskip
\noindent
{\bf ACKNOWLEDGMENTS}
\bigskip

This research is supported by the Carmel Science Foundation. 
I would like to thank Yael Oren, Arbel M. Ongo, Ayelet B. Lata, and Alona B. 
Tea for helpful discussions.



\begin{thebibliography}{99}

\bibitem{Bar} J. M. Bardeen, W. H. Press and S. A. Teukolsky, Astrophys. J. {\bf 178}, 347 (1972).

\bibitem{Chan} S. Chandrasekhar, {\it The Mathematical Theory of Black
Holes}, (Oxford University Press, New York, 1983).

\bibitem{ShTe} S. L. Shapiro and S. A. Teukolsky, {\it Black Holes, White
Dwarfs and Neutron Stars: The Physics of Compact Objects}, 1st ed.
(Wiley-Interscience, 1983).

\bibitem{Pod} M. A. Podurets, Astr. Zh. {\bf 41}, 1090 (1964) [English translation in
Sovet Astr.-AJ {\bf 8}, 868 (1965)].

\bibitem{Ame} W. L. Ames and K. S. Thorne, Astrophys. J. {\bf 151}, 659 (1968).

\bibitem{Ste} I. Z. Stefanov, S. S. Yazadjiev, and G. G. Gyulchev, Phys. Rev.
Lett. {\bf 104}, 251103 (2010).

\bibitem{Goe} C. J. Goebel, Astrophys. J. {\bf 172}, L95 (1972).

\bibitem{Mas} B. Mashhoon, Phys. Rev. D {\bf 31}, 290 (1985).

\bibitem{Dol} S. R. Dolan, Phys. Rev. D {\bf 82}, 104003 (2010).

\bibitem{Dec} Y. De\'canini, A. Folacci, and B. Raffaelli, Phys. Rev. D {\bf 81},
104039 (2010); Y. De\'canini, A. Folacci, and B. Raffaelli, Phys.
Rev. D {\bf 84}, 084035 (2011).

\bibitem{Hodf} S. Hod, Phys. Rev. D {\bf 80}, 064004 (2009)
[arXiv:0909.0314]; S. Hod, Phys. Rev. D {\bf 78}, 084035 (2008)
[arXiv:0811.3806]; S. Hod, Phys. Rev. D {\bf 75}, 064013 (2007)
[arXiv:gr-qc/0611004]; S. Hod, Class. Quant. Grav. {\bf 24}, 4235
(2007) [arXiv:0705.2306]; S. Hod, Phys. Lett. B {\bf 715}, 348
(2012) [arXiv:1207.5282].

\bibitem{Hodt1} S. Hod, Phys. Rev. D {\bf 84}, 124030 (2011)
[arXiv:1112.3286]; S. Hod, Phys. Rev. D {\bf 84}, 104024 (2011)
[arXiv:1201.0068].

\bibitem{Hodt2} S. Hod, Phys. Lett. B {\bf 718}, 1552 (2013) [arXiv:1210.2486]; S.
Hod, Phys. Lett. B {\bf 751}, 177 (2015) [arXiv:1707.06246]; S. Hod,
Class. Quant. Grav. {\bf 33}, 114001 (2016) [arXiv:1705.08905].

\bibitem{Zde1} J. Novotn\'y, J. Hlad\'ik, and Z. Stuchl\'ik, Phys. Rev. D {\bf 95}, 043009 (2017).

\bibitem{Hodcom} S. Hod, Phys. Rev. D {\bf 97}, 084018 (2018) [arXiv:1810.03618].

\bibitem{IIKPR} A. Ianniccari, A. J. Iovino, A. Kehagias, D. Perrone, and A. Riotto, 
Phys. Rev. Lett. {\bf 133}, 081401 (2024) [arXiv:2404.02801]. 

\bibitem{Chop1} M. W. Choptuik, Phys. Rev. Lett. {\bf 70}, 9 (1993).

\bibitem{HodPir1} S. Hod and T. Piran, Phys. Rev. D {\bf 55}, 3485 (1997) [arXiv:gr-qc/9606093]; 
S. Hod, M. Sc. thesis, The Hebrew University, Jerusalem, (1995).

\bibitem{HodPir2} S. Hod and T. Piran, Phys. Rev. D {\bf 55}, R440 (1997) [arXiv:gr-qc/9606087]. 

\bibitem{Noteunits} We use natural units in which $G=c=\hbar=1$.

\bibitem{Notepm} Here a prime $'$ denotes a differentiation with respect to the radial coordinate $r$.

\bibitem{Noterho} Here $\rho=-T^t_t$ is the enrgy density of the self-gravitating matter configuration. 


\bibitem{Hodev1} S. Hod, Phys. Lett. B {\bf 739}, 383 (2014) [arXiv:1412.3808]. 

\bibitem{Cun} P. V. P. Cunha, E. Berti, and C. A. R. Herdeiro, Phys. Rev. Lett. {\bf 119}, 251102 (2017).

\bibitem{Hodev2} S. Hod, Phys. Lett. B {\bf 776}, 1 (2018) [arXiv:1710.00836].

\bibitem{Noteaaa} Here we have used the relations $a'={\cal C'}/(1-2{\cal C})^{3/2}$ 
and $a''=3{\cal C}'^2/(1-2{\cal C})^{5/2}+{\cal C}''/(1-2{\cal C})^{3/2}$ 
[see Eqs. (\ref{Eq4}) and (\ref{Eq6})]. 

\bibitem{Notebrd} Here we assume that the compactness function is broad enough. 
 
\end{thebibliography}
\end{document}